\documentclass[10pt,journal,compsoc]{IEEEtran}%
\usepackage{cite}
\usepackage{amsmath,amssymb,amsfonts}
\usepackage{algorithmic}
\usepackage{graphicx}
\usepackage{textcomp}
\usepackage{hyperref}
\usepackage{algorithmic}
\usepackage{algorithm}
\usepackage[table,xcdraw]{xcolor}
\usepackage{comment}
\usepackage{titlesec}
\usepackage[table,xcdraw]{xcolor}
\usepackage[font=footnotesize,skip=0pt]{caption}
\titlespacing{\subsection}{0pt}{*0.6}{*0.5}
\def\BibTeX{{\rm B\kern-.05em{\sc i\kern-.025em b}\kern-.08em
    T\kern-.1667em\lower.7ex\hbox{E}\kern-.125emX}}
\begin{document}
\renewcommand{\paragraph}[1]{\textbf{#1}\mbox{}\\}
\newcommand{\sharif}[1]{\textsf{\color{blue}{[{#1 -- Sharif}]}}}


\title{Privacy Preserving Charge Location Prediction for Electric Vehicles \\
}


\author{%
\IEEEauthorblockN{\IEEEauthorrefmark{1}Robert Marlin, \IEEEauthorrefmark{1}Raja Jurdak, Alsharif Abuadbba\IEEEauthorrefmark{2},  \IEEEauthorrefmark{1}Dimity Miller} \\
%

\IEEEcompsocitemizethanks{\IEEEcompsocthanksitem \IEEEauthorrefmark{1}Robert Marlin, \IEEEauthorrefmark{1}Raja Jurdak with School of Computer Science and \IEEEauthorrefmark{1}Dimity Miller with School of Electrical Engineering and Robotics, Queensland University of Technology, Australia. \IEEEauthorrefmark{1}Robert Marlin is also with CSIRO's Data61 and Cyber Security Cooperative Research Centre, Australia\{robert.marlin;r.jurdak;d24.miller\}@qut.edu.au.}
\IEEEcompsocitemizethanks{\IEEEcompsocthanksitem \IEEEauthorrefmark{2}Alsharif Abuadbba is with CSIRO's Data61, Australia\{sharif.abuadbba\}@data61.csiro.au.}
}

\markboth{Privacy Preserving Charge Location Prediction for Electric Vehicles, November~2024}%
{Shell \MakeLowercase{\textit{et al.}}: Bare Demo of IEEEtran.cls for Computer Society Journals}

\maketitle
\begin{abstract}
By 2050, electric vehicles (EVs) are projected to account for 70\% of global vehicle sales. While EVs provide environmental benefits, they also pose challenges for energy generation, grid infrastructure, and data privacy. Current research on EV routing and charge management often overlooks privacy when predicting energy demands, leaving sensitive mobility data vulnerable. To address this, we developed a Federated Learning Transformer Network (FLTN) to predict EVs' next charge location with enhanced privacy measures. Each EV operates as a client, training an onboard FLTN model that shares only model weights, not raw data with a community based Distributed Energy Resource Management System (DERMS), which aggregates them into a community global model. To further enhance privacy, non-transitory EVs use peer-to-peer weight sharing and augmentation within their community, obfuscating individual contributions and improving model accuracy. Community DERMS global model weights are then redistributed to EVs for continuous training. Our FLTN approach achieved up to 92\% accuracy while preserving data privacy, compared to our baseline centralised model, which achieved 98\% accuracy with no data privacy. Simulations conducted across diverse charge levels confirm the FLTN’s ability to forecast energy demands over extended periods. We present a privacy focused solution for forecasting EV charge location prediction, effectively mitigating data leakage risks.
\end{abstract}



\section {\textbf{Introduction}} \label{sec:Introduction}

\par The global shift from internal combustion engine vehicles (ICEVs) to Electric Vehicles (EVs) marks a critical move toward sustainable transportation, with EVs projected to account for up to 70\% of global vehicle sales by 2050 \cite{EnerOutlook21}. While this transition supports environmental goals, it also increases energy demands within communities and cities. Managing this demand across large EV networks will require advanced technology, with Distributed Energy Resource Management Systems (DERMS) playing a crucial role in enabling efficient energy management. However, processing EV data through DERMS raises critical privacy and security concerns due to the sensitivity of location history, charging patterns, and trip details. Federated Learning (FL) has emerged as a promising method to address these concerns by decentralising model training, allowing data to remain on devices rather than being transmitted to a central server \cite{mcmahan2017communication}. Traditional FL models are generally designed for static environments with relatively stable client data distributions, making them less effective for highly mobile agents such as EVs. The lack of focus on leveraging mobility for privacy leaves a gap in existing FL approaches, particularly for dynamic, geographically dispersed data sources like EV networks. To address these concerns, our FLTN solution embraces agent mobility for privacy through a mechanism called peer weight sharing, and enables compliance with data privacy regulations, such as the European Union's GDPR \cite{regulation2018general} and California's CCPA \cite{de2018guide}.

\par The FLTN framework enhances data privacy by decentralising model training, addressing the challenge of managing sensitive, distributed data across large, dynamic EV networks. To achieve this, community DERMS collect local model weights from EVs during charging sessions. For non-transitory EVs, this process includes peer-to-peer weight sharing and augmentation, further strengthening privacy measures. Augmentation, in the context of this work, involves non-transitory EVs collaboratively combining their locally trained model weights through a peer-to-peer sharing mechanism before sending the combined weights to the community DERMS. This obfuscates the origin of individual weight updates, breaking the direct link between an EV's data and its contribution to the global model. Additionally, augmentation mitigates the influence of outliers, enhancing the robustness and reliability of the FL process while maintaining convergence efficiency. Achieving reasonable predictive accuracy in this framework typically requires a minimum of 50–100 EVs. As more EVs contribute, the diversity and representativeness of the data improve, further boosting the model’s performance. Using Federated Averaging (FedAVG), DERMS aggregates these weights, including those from transitory EVs, to form a community global model. FedAVG is well-suited to bandwidth-limited environments, such as EVs with constrained processing power \cite{mcmahan2017communication}, and effectively manages diverse non-IID data patterns across EVs, like varying charging habits and usage behaviors, enhancing accuracy while preserving privacy. FLTN further mitigates risks from malicious EVs that could upload falsified updates by exchanging model weights rather than gradients, which are more vulnerable to inversion and backdoor attacks \cite{duan2021ssgd}. Additionally, DERMS modeling weights across diverse EVs dilutes any single EV’s influence, reducing the impact of injected false data on the community model.

\par EV mobility strengthens privacy, as frequent movement across community DERMS naturally mixes data contributions. However, non-transitory EVs that remain stationary within regions face higher privacy risks. Our approach introduces peer-to-peer weight sharing and augmentation as an additional defense for these EVs, reducing the exposure of individual model updates. Techniques like batch normalization and large batch sizes further obscure specific data points within aggregated updates, enhancing privacy. FLTN’s decentralised, privacy-focused framework not only improves model accuracy but also meets legislative privacy requirements. In summary, this paper’s contributions are as follows:
\vspace{1mm} 

\begin{enumerate}
\item We introduce a localised community FLTN system capable of predicting EVs' next charge locations while preserving user data privacy. By sharing only model weights with a community DERMS server, private data leakage is prevented. This novel application of FLTN to mobile EV agents for charge location prediction offers a scalable, privacy-enhancing solution that maintains high prediction accuracy across diverse, resource-constrained clients.
\item For non-transitory EVs, we ensure data privacy through peer-to-peer model weight sharing and augmentation, obfuscating individual identities within each community. Our FLTN-based mobility prediction model also leverages large batch sizes and normalization to further protect user data against malicious actors, while maintaining high predictive accuracy.
\item We evaluate our FLTN model using a novel EV mobility dataset. A key challenge is achieving high prediction accuracy while preserving privacy across decentralized networks. Our results show that accurate predictions require a minimum of 100 to 150 EVs per community, achieving an accuracy of 92\%. Additionally, the model can predict the next charge location up to three days in advance, balancing accuracy and privacy for scalable, real-world applications.
\end{enumerate}

\par The remainder of this paper is organised as follows. Section~\ref{sec: Related Work} details related works in privacy preservation for mobility data, federated learning for data privacy, and malicious actors within mobility domains. Section~\ref{sec: Problem Definition} formalises our FLTN EV next charge prediction problem. Section~\ref{sec: Mathematical Framework} introduces the model's architecture and optimisation process. Section~\ref{sec: Proposed Solution} discusses our approach to privacy-preserving EV next charge location prediction, this includes our approach to our novel EV taxi mobility dataset as well as our proposed solution for privacy-preserving EV next charge location prediction. Section~\ref{sec: Performance Evaluation} evaluates our proposed distributed approach to this problem area. Section~\ref{sec: Discussion} discusses any possible trade-offs, lastly we conclude our paper with a discussion on future work. 


\section {\textbf{Related Work}} \label{sec: Related Work}

\subsection {\textbf {Privacy Preservation for Mobility Data}}

\par Mobility data, which encompasses the tracking and analysis of transportation modes, is increasingly essential for applications like urban planning, traffic management, and personalised services. However, its misuse can lead to privacy breaches, exposing individuals to unwanted surveillance, discrimination, or identity theft \cite{allahrakha2023balancing}. Preserving privacy in mobility data is challenging due to its sensitive nature, often including granular details such as visited locations, time spent, and travel routes. As Miguel et al. \cite{andres2013geo} highlight, even anonymised data can be re-identified through linkage with external datasets, a process known as de-anonymisation.

\subsubsection {\textbf {Techniques for Privacy Preservation}}

\par Various techniques have been proposed to address privacy challenges in mobility data. Differential privacy is widely used, introducing noise to data to obscure true values while allowing aggregate analysis, ensuring that query outputs are statistically indistinguishable regardless of an individual's presence \cite{gruteser2003anonymous}. Building on these principles, our solution enhances privacy by employing peer-to-peer model weight sharing and augmentation among locally stationed (non-transitory) EVs before community DERMS modeling. During this step, non-transitory EVs exchange their locally trained model weights with peers in the same community, combining them into a shared set of augmented weights. This process obfuscates the origin of individual EV updates, reducing the risk of direct exposure to sensitive data. Additionally, by incorporating contributions from multiple EVs, this step mitigates the influence of potential outliers, enhancing the robustness of the combined updates. The approach assumes that neighboring EVs act as 'honest but curious' participants, acknowledging that a malicious EV could theoretically infer patterns from shared weights, though this risk is reduced by the obfuscation introduced through augmentation. However, augmenting weights within peer groups before transmission to the community DERMS helps obscure individual contributions, aligning with the principles of differential privacy by making specific data points harder to isolate. Unlike traditional differential privacy techniques, which typically inject noise into the data or model updates and may reduce prediction accuracy, our FLTN framework achieves strong privacy without sacrificing data quality. This is accomplished by directly sharing and augmenting model weights among peers, preserving both privacy and predictive performance.

\subsubsection {\textbf {Federated Learning for Data Privacy}}

\par Recently, FL has emerged as a privacy-preserving approach by training models across devices without data sharing, offering privacy advantages over differential privacy methods \cite{sweeney2002k}. One such study applied FL in EV charging contexts, demonstrating FL’s potential but also highlighting specific limitations that our approach aims to address. For instance, 'Predicting Electric Vehicle Charging Stations Occupancy: A Federated Deep Learning Framework' proposes an FL framework for predicting charging station occupancy. While it achieves accurate occupancy forecasts, it is limited to station-specific predictions and does not generalise to individual EVs’ next charge locations, which our approach targets \cite{douaidi2023predicting}. In another study using FL by Teimoori et al.,\cite{teimoori2022secure} a secure, cloudlet-based recommendation system is introduced, focusing on recommending stations rather than predicting future charging locations. This system is effective for recommendations but does not address the need for location privacy in direct charge location predictions. Another study by Li et al., \cite{li2024federated} is primarily centered around station demand and is not designed to forecast individualised EV next charge locations, a gap that our FLTN approach fills.  

\par Traditional methods like differential privacy and k-anonymity, while effective, have notable limitations: differential privacy reduces accuracy due to noise, and k-anonymity sacrifices utility through data suppression. FL models that share gradients are also vulnerable to inversion attacks. Our FLTN approach addresses these limitations by sharing model weights instead of gradients, mitigating inversion risks, and preserving accuracy without noise. The use of large batch sizes and normalisation techniques improves model stability, while peer-to-peer weight sharing and augmentation among non-transitory EVs before DERMS updates further conceals individual data contributions. Additionally, the inherent mobility of transitory EVs disperses data contributions across communities, enhancing privacy by diluting individual data traces. 

\subsubsection {\textbf {Swarm Learning Approach for Data Privacy}}

\par Swarm learning, a decentralised machine learning approach, has gained attention as an alternative to traditional FL. Unlike FL, which typically relies on a central aggregator, swarm learning enables participants to share and update model weights directly in a peer-to-peer fashion, often using a ring or daisy-chain topology \cite{warnat2021swarm}. This architecture reduces dependence on a central server, enhancing resilience and privacy. However, the sequential nature of weight exchanges in swarm learning can introduce increased communication latency and risks of propagating corrupted updates through the network.

\par Our approach draws inspiration from swarm learning’s decentralised principles but adapts them to better accommodate the dynamic mobility patterns of EVs. Specifically, non-transitory EVs in the FLTN framework augment weights by combining contributions from all peers within a community before sharing with the DERMS. A key aspect of this process is the assignment of alphas, which determine the relative weight of each EV’s contribution during augmentation. In this work, we assume equal alphas across all non-transitory EVs, ensuring uniform contributions and simplifying the augmentation mechanism. Unlike the sequential weight-sharing in swarm learning, this parallel peer-to-peer augmentation obfuscates the origin of individual updates while preserving robust model convergence. Additionally, the FLTN framework incorporates mechanisms tailored to EV spatio-temporal characteristics, enhancing predictive accuracy and scalability compared to existing swarm learning techniques.\\

\paragraph{\textbf{Comparison to Well-Known Aggregation Functions}}

\par Our augmentation function extends traditional federated aggregation methods like FedAvg by introducing a peer-to-peer pre-combination step. Unlike FedAvg, where individual updates are directly aggregated at the central server, the augmentation strategy obfuscates the origin of updates by combining contributions within a peer group before submission. This enhances privacy by reducing the risk of associating individual EV updates with specific data points. Compared to swarm learning, which relies on sequential weight-sharing, the augmentation function operates in parallel, reducing communication latency and improving scalability. While it lacks the outlier detection features of advanced aggregation methods like Krum \cite{zhang2022dim}, the augmentation process inherently mitigates the impact of outliers by combining updates locally, offering a simpler yet effective mechanism for robust and privacy-preserving FL.

\subsubsection {\textbf {Attacks in Edge Machine Learning}} 
\par Data poisoning attacks pose a serious threat in edge machine learning, where models are trained directly on source-collected data. Adversaries exploit the decentralised structure and data variability of edge computing to manipulate training data and degrade model performance \cite{zhou2022differentially}. FLTN mitigates this threat through several mechanisms. By sharing model weights rather than raw data or gradients, FLTN limits the impact of any single EV’s contribution, reducing the likelihood that a malicious actor can substantially influence the DERMS community model. The peer-to-peer weight sharing and augmentation process for non-transitory EVs further dilutes individual contributions before updates reach the community model, making it more challenging for injected false data to significantly alter model outcomes. Additionally, the natural agent mixing resulting from EV mobility enhances privacy and mitigates poisoning risk by distributing data contributions across different communities. 

\par Model inversion attacks also pose a significant threat to the privacy and security of machine learning models. This vulnerability arises from the overexposure of model outputs, which adversaries can exploit to gain insights into the underlying data distribution, effectively revealing private information through repeated interactions \cite{zhu2019deep}. The risk is heightened in edge computing environments, where models interact with potentially untrusted users, as is common in federated systems \cite{li2024survey}. Our FLTN framework mitigates these inversion risks by sharing only model weights rather than gradients or raw data which significantly reduces the exposure of sensitive data while maintaining data utility. By sharing only the final trained weights from each local model, rather than gradients or activations, we minimise the potential for inversion attacks. Gradients, which represent the direction of model updates during backpropagation, can reveal sensitive information if accessed by a malicious actor. Activations are also excluded, as they could be used to infer input data characteristics, thus protecting sensitive user data.

\par While much of the existing literature addresses vulnerabilities in EV mobility systems through regulatory frameworks or broad technological approaches, this paper introduces a novel FLTN framework specifically designed for EV charge location prediction with enhanced privacy measures. Unlike previous work reliant on centralised data or techniques like differential privacy, our method leverages decentralised model training with a multi-layered privacy approach. By combining peer-to-peer model weight sharing and augmentation, large batch training, and aggregation of model weights rather than gradients, our FLTN framework mitigates common privacy risks while preserving accuracy. The inherent mobility of EVs further enhances privacy by dispersing data contributions across communities, naturally obfuscating individual data traces. This integrated approach significantly reduces the risks of data leakage and model inversion attacks, offering a robust, privacy-preserving solution to evolving threats in the mobility domain.


\section {\textbf {Problem Definition}} \label{sec: Problem Definition}

\par This section outlines the problem addressed in this paper: predicting the next location where an EV will recharge its battery. We introduce the relevant background concepts and notation necessary for understanding the EV mobility dataset, particularly focusing on the prediction task within the context of protecting EV data privacy using our FLTN system. Unlike traditional methods that rely on centralised data repositories, our approach safeguards EV data privacy by sharing only locally trained model weights with a community-based DERMS. Our study considers the spatio-temporal mobility patterns associated with EV-based services, including features such as pick-up locations, adjacent community areas, current battery levels, and timestamps for each trip. Understanding these factors is crucial for accurately predicting where an EV is likely to recharge next.

\par Our goal is to develop a model capable of predicting an EV's next charging location, represented by the community area where the EV is expected to be when its battery reaches a critical charging threshold. Community areas are defined as distinct regions within or around a city, acting as potential sites for pick-ups, drop-offs, or EV charging. To ensure data privacy, we use an FLTN system that allows individual EVs to train locally on their respective datasets. EVs share only local model weights with community-based DERMS, which act as distributed servers that train on EV (client) weights. For locally stationed (non-transitory) EVs, peer-to-peer model weight sharing and augmentation occur before submitting weights to DERMS, adding an extra layer of privacy. Community DERMS then share their aggregated global model weights with EVs during charging sessions. This approach mitigates the risk of exposing sensitive data containing locations, timestamps, and social interactions while enabling effective prediction of the next charging location.
\vspace{3mm}
\hrule\vspace{1mm}
Where:
\vspace{1mm}\hrule\vspace{1mm}
\begin{itemize}
    \item \(\mathbf{W}_i^{(t)}\): The model weights on local client \(i\) (an EV) at training iteration \(t\).
    \item \(N\): The total number of EVs (local clients) participating in the federated learning process.
    \item \(\text{DERMS}(\cdot)\): The distributed energy resource management system (DERMS) aggregates local model weights from EVs and updates the community global model weights.
    \item \(\mathbf{W}^{(t+1)}\): The updated community global model weights after aggregation by DERMS at iteration \(t+1\).
    \item \(\hat{y}_{\text{EV}}\): The predicted next charge location for the EV.
    \item \(\text{FL\_Transformer}(\cdot)\): The federated learning transformer model that takes input data and the community model weights to predict the next charging location.
    \item \(\mathbf{X}_{\text{EV}}\): The input feature set for the EV, including pick-up locations, neighbouring community areas, current battery level, time stamps, etc.
    \item \(\text{PeerShare}(\cdot)\): The peer-to-peer sharing and augmentation of local model weights among non-transitory EVs before sharing with DERMS.
\end{itemize}
\vspace{1mm}\hrule\vspace{1mm}

\section*{Federated Learning Equation for Predicting EV Next Charge Location}

\textbf{Global Model Weight Update (with Peer-to-Peer Sharing and Augmentation):}
\begin{equation}
\mathbf{W}^{(t+1)} = \text{DERMS}\left(\frac{1}{N} \sum_{i=1}^{N} \text{PeerShare}(\mathbf{W}_i^{(t)}) \right)
\end{equation}
\textbf{Prediction Equation:}
\begin{equation}
\hat{y}_{\text{EV}} = \text{FL\_Transformer}(\mathbf{X}_{\text{EV}}, \mathbf{W}^{(t+1)})
\end{equation}

\section*{Explanation}

\begin{enumerate}
    \item \textbf{Local Training:} Each EV \(i\) trains a local FL transformer model using its data to update its model weights \(\mathbf{W}_i^{(t)}\).
    
    \item \textbf{Peer-to-Peer Sharing and Augmentation:} For non-transitory EVs, local model weights are shared within peer groups, and the weights are augmented using \(\text{PeerShare}(\cdot)\) before being sent to DERMS.
    
    \item \textbf{Weight Aggregation:} The DERMS servers, distributed across each community, aggregate the weights from all participating EVs (including augmented peer weights) to form a community model \(\mathbf{W}^{(t+1)}\).
    
    \item \textbf{Community Model Update:} The aggregated community weights \(\mathbf{W}^{(t+1)}\) are shared with the EVs during charging sessions. Each EV then uses these updated weights for future training of their local on-board model.
    
    \item \textbf{Prediction:} Using the community model weights \(\mathbf{W}^{(t+1)}\), the FL transformer model on each EV predicts the next charging location \(\hat{y}_{\text{EV}}\).
\end{enumerate}
\hrule
\vspace{2mm} 

This equation represents the iterative process of federated learning where local models contribute to a community global model without sharing raw data, thereby preserving privacy while still enabling accurate predictions.


\section{Mathematical Framework} \label{sec: Mathematical Framework}

\subsection{Local Model Training and Peer-to-Peer Sharing with Augmentation}
Let \(\mathcal{D}_i\) represent the local dataset for EV \(i\), where \(i \in \{1, 2, \dots, N\}\) and \(N\) is the total number of EVs. Each EV \(i\) trains a local model \(f_i(W_i^t)\) on its dataset \(\mathcal{D}_i\), where \(W_i^t\) represents the model parameters (weights) for EV \(i\) at time \(t\). The local objective function to minimise is given by:
\begin{equation}
    W_i^t = \arg \min_{W} \mathcal{L}_i(f_i(W), \mathcal{D}_i)
\end{equation}
where \(\mathcal{L}_i\) is the loss function for EV \(i\).
After local training, non-transitory EVs engage in peer-to-peer sharing and augmentation. Each EV \(i\) shares its local weights \(W_i^t\) with its peer group, receiving peer weights \(W_{pg}^t\) from other non-transitory EVs. These peer weights are then used to augment the EV's own weights, introducing diversity and obfuscating individual EV contributions. The augmented weights are defined as:
\begin{equation}
    W_{aug}^t = W_i^t + \alpha \sum_{j \in \text{peers}} W_j^t
\end{equation}
where \(\alpha\) is a scaling factor controlling the contribution of peer weights. The augmented weights \(W_{aug}^t\) are then transmitted to the community DERMS. In contrast, transitory EVs send their locally trained weights \(W_i^t\) directly to the DERMS without peer-to-peer sharing or augmentation.
\vspace{2mm}
\subsection{Community Model Aggregation by DERMS}
Each community DERMS aggregates the model weights from all participating EVs, including the augmented weights from non-transitory EVs, to update the community global model. The community model parameters \(\Theta\) are computed as:
\begin{equation}
    \Theta = \frac{1}{N} \sum_{i=1}^{N} \tilde{W}_i^t
\end{equation}
where \(\tilde{W}_i^t\) represents either the augmented weights \(W_{aug}^t\) for non-transitory EVs or the original weights \(W_i^t\) for transitory EVs.
\vspace{2mm}
\subsection{Model Update During Charging Transaction}
When an EV \(i\) connects to a charging station, it receives the latest community global model weights \(\Theta\) from the DERMS. The EV then updates its local model as follows:
\begin{equation}
    W_i^{t+1} \leftarrow \Theta
\end{equation}
The updated local model \(f_i(W_i^{t+1})\) is now ready for further training.
\vspace{2mm}
\subsection{Federated Learning Process}
The overall federated learning process, including peer-to-peer sharing and augmentation, can be summarised as follows:

\begin{algorithm}[H]
\caption{Federated Learning Process for EVs with DERMS and Peer-to-Peer Sharing and Augmentation}
\begin{algorithmic}[1]
\STATE Initialise community model weights \(\Theta\)
\FOR {each round \(t = 1, 2, \dots\)}
    \FOR {each EV \(i \in \{1, \dots, N\}\) \textbf{in parallel}}
        \STATE Train local model \(f_i(W_i^t)\) on \(\mathcal{D}_i\)
        \IF{EV \(i\) is non-transitory}
            \STATE Perform peer-to-peer sharing: \(W_{pg}^t = \text{PeerShare}(W_i^t)\)
            \STATE Augment local weights with peer weights: \(W_{aug}^t = W_i^t + \alpha \sum_{j \in \text{peers}} W_j^t\)
            \STATE Send augmented weights \(W_{aug}^t\) to DERMS
        \ELSE
            \STATE Send local model weights \(W_i^t\) to DERMS
        \ENDIF
    \ENDFOR
    \STATE Aggregate community global model weights: \(\Theta = \frac{1}{N} \sum_{i=1}^{N} \tilde{W}_i^t\)
    \FOR {each EV \(i \in \{1, \dots, N\}\)}
        \STATE During charge transaction, update local model: \(W_i^{t+1} \leftarrow \Theta\)
    \ENDFOR
\ENDFOR
\end{algorithmic}
\end{algorithm}

\par This paper presents our FLTN system where EVs train local transformer models and share their model weights with a community DERMS. For non-transitory EVs, peer-to-peer sharing and augmentation are performed before distributing weights to the community DERMS, enhancing model robustness and privacy. The framework allows for distributed model training without sharing raw data, preserving data privacy while enabling collective model improvements.


\section {\textbf{Proposed Solution}}\label{sec: Proposed Solution}

\par This section discusses three fundamental components of our research: 1) the solution architecture, 2) our taxi EV mobility dataset, and 3) the proposed FLTN system to secure private EV data. First, we present the system pipeline, outlining the sequence of processes from data input to the final prediction of EV's next charge location. We detail the key stages, including data preprocessing, and the prediction mechanism, emphasising the role of each in the overall system. Second, we describe the construction and characteristics of the taxi EV mobility dataset, highlighting the methodologies used for data collection, feature engineering, and the assumptions incorporated into the dataset. Last, we explore the implementation of our FLTN system, explaining its structure, training process, and its specific application in predicting the EV's next charging location.
\vspace{5mm}
\subsection {\textbf{Proposed Pipeline}}\label{sec: Pipeline_Diagram}

\begin{figure*}[h!]
\centering
\includegraphics[width=15.5cm, height=5.5cm]{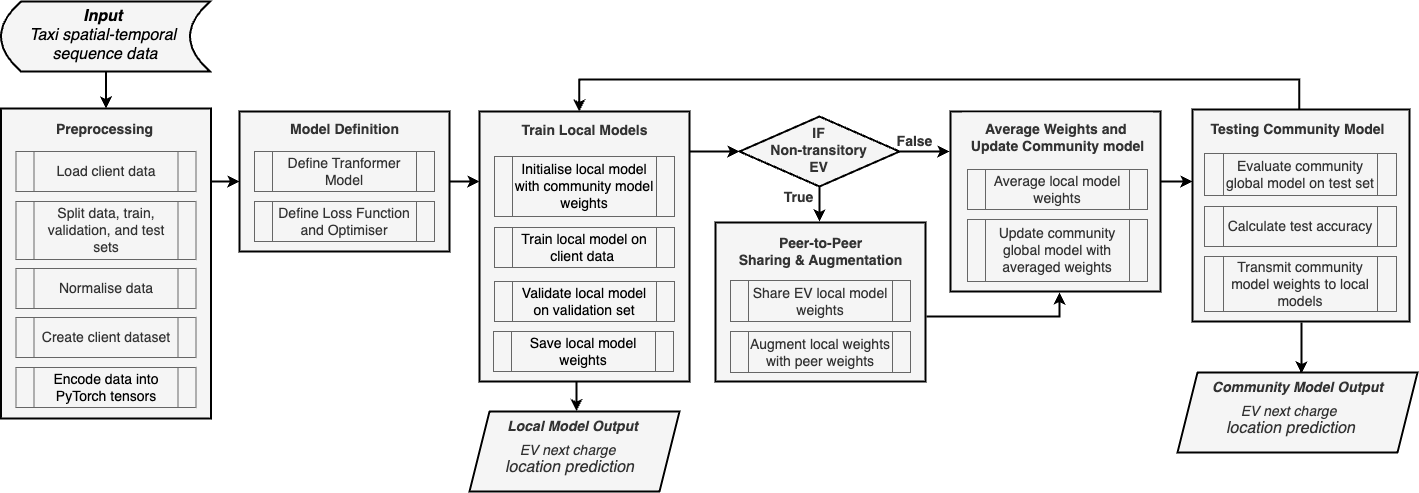}
\vspace{4mm}
\caption{The pipeline diagram above consists of processes and stages for our FLTN system from data input to our final output EV next charge location prediction.}
\label{fig: Pipeline_Diagram}
\end{figure*}

\par Our pipeline architecture, shown in Figure~\ref{fig: Pipeline_Diagram}, summaries processes within our proposed FLTN solution. This includes data loading and pre-processing, model definition, training local models, peer-to-peer sharing and augmentation, averaging weights and updating both our community global model and our local model, testing our community model, and finally, prediction output for the global and local model being the EV next charge location prediction. This process enables decentralised training across multiple EV clients. Each EV dataset, containing spatio-temporal sequences is modelled locally to ensure data privacy. The local models on each client are trained independently. 

\par For non-transitory EVs, local model weights are shared directly with other non-transitory EVs within the community in a peer-to-peer manner. In this process, each EV receives model weights from its peers and combines them with its own weights during augmentation. This step incorporates diverse data patterns from nearby EVs, enhancing the robustness of the local model while obfuscating individual EV characteristics. The augmented weights are then transmitted to the local community DERMS, where they contribute to the community model, further preserving privacy by masking individual EV-specific information. This augmentation strategy naturally increases the relative importance of non-transitory EVs in the DERMS global model, as their updates represent combined contributions from multiple peers within the community. In contrast, transitory EVs share their local model weights directly with the community DERMS. These updates are essential for capturing the dynamic mobility patterns of EVs, contributing diversity and generalisability to the model. Together, the contributions of non-transitory and transitory EVs strike a balance between local stability and global adaptability, ensuring robust and accurate predictions. The community DERMS aggregates these weights using FedAVG to create a community global model, which is subsequently redistributed to all EV clients. This architecture promotes privacy preservation while enabling effective learning from the non-IID datasets inherent in EV mobility patterns.

\vspace{5mm}

\subsection {\textbf{EV Taxi Dataset}}\label{sec: Data} 

\par Our proposed solution, leveraging machine learning, required a large dataset containing EV mobility and charge transactions. For this, we used a dataset created in \cite{marlin2024electric}, comprising both empirical and synthetic data. The empirical component is based on a real-world Chicago taxi mobility dataset (non-EV taxis), while the synthetic data incorporates EV industry metrics \cite{chicago2016}. This dataset assumes that (1) most taxis will be EVs by 2050, and (2) all charge transactions will use uniform charge point technology and speeds. EVs in the dataset adhere to industry standards for nine EV models relevant to taxi services, with battery capacities ranging from 143 km to 416 km. Taxi assignments to EV models were randomised.

\begin{figure}
    \centering 
    \includegraphics[width=7cm, height=8cm]{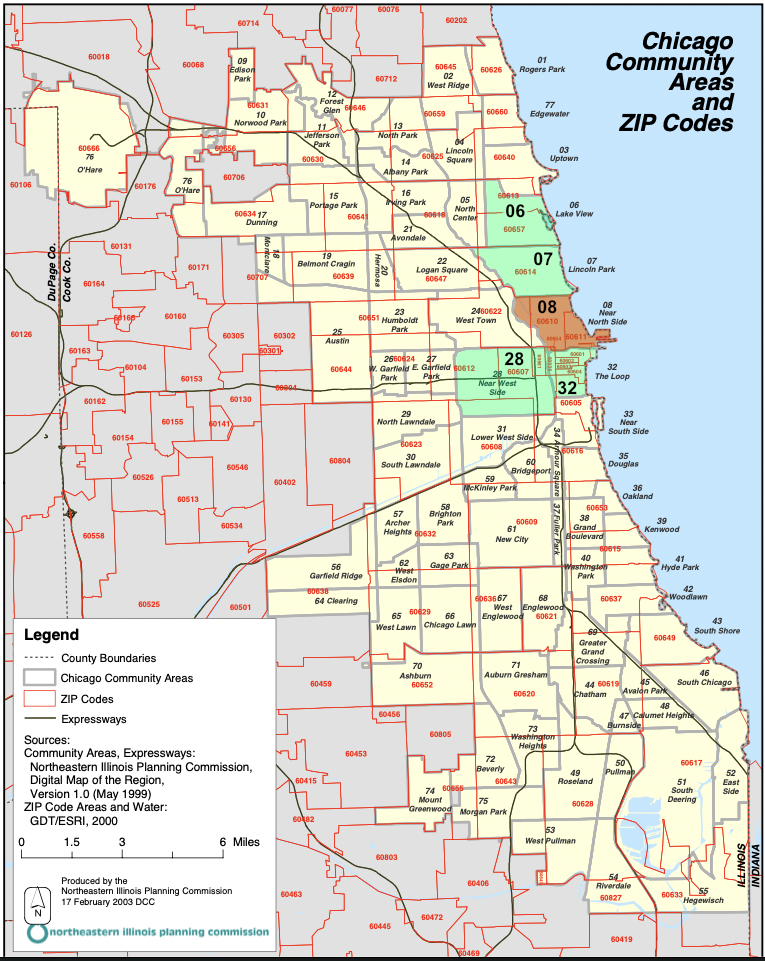}
    \vspace{4mm}
    \caption{Chicago city map - community areas light yellow. The top 5 communities for taxi activity in green and orange are 32, 6, 28, 7, and 8 which has the highest density of taxi activity in orange.}
    \label{fig: Top 5 community areas, 32, 6, 28, 7, and 8}
\end{figure}

\par Other important dataset features include 77 community areas within the city of Chicago as displayed in Figure \ref{fig: Top 5 community areas, 32, 6, 28, 7, and 8}, each of these community areas representing either possible pick-up, drop-off, or charge locations. Each EV within the dataset represents a client in the FLTN system. All EV taxis vary in their number of trips. Taxi trips span a 12-month period (one year) which represents the temporal scope of this study. This dataset includes a range of battery charge levels from 100\% battery charge level to 20\% battery charge level. This range of battery charge levels allows our model to understand a variety of charge periods where our model can forecast over multiple time spans from the near-immediate to periods of days into the future. We select 20\% battery charge level as a baseline to charge following industry standards for EV battery health \cite{batteryhealth}. 

\vspace{5mm}
\subsection {\textbf{Our Proposed Solution}}\label{sec: Modelling}

\par This section first discusses FL for data privacy. We then discuss model selection for our proposed solution. Following this we discuss training, validation, test outcomes, and methods. Then we look at performance evaluation, followed by baseline modelling and finally, we cover results and analysis. 

\subsubsection {\textbf {Federated Learning for Data Privacy}}  \label{sec: Federated Learning}

\par In a previous study \cite{marlin2024electric}, we used a centralised CNN model to predict EVs' next charge location. While this provided accurate predictions, it lacked data privacy considerations. To address this, we employ FL, a decentralised approach that preserves data privacy by enabling local model training on EV mobility spatio-temporal data \cite{ji2024emerging}. However, applying FL in this context is not straightforward; achieving reliable accuracy depends on aggregating contributions from a sufficient number of EVs, which varies based on community size and mobility patterns. Determining the optimal number of EVs is crucial for effective learning, as sparse participation may lead to reduced predictive accuracy and data utility. Data leakage risks, including insights into locations, mobility patterns, and transaction behaviors, are reduced as FL allows each EV to train locally, sharing only model updates. In our implementation, EVs share local model weights with a community-based DERMS, which aggregates these into a community global model across 77 communities in the city of Chicago. This decentralised approach ensures raw data remains on vehicles, significantly reducing privacy risks and meeting data protection standards \cite{bakare2024data}.

\par The use of a decentralised FL network specifically for our study also addresses challenges for data heterogeneity and the need for robust, generalised models across different types of vehicles and driving environments \cite{wu2024application}. However, implementing FL in EVs is challenging. Security threats such as model inversion attacks within FL networks pose significant privacy risks by exploiting the shared model updates to reconstruct sensitive information from the training data \cite{issa2024rve}. In such an attack, an adversary with access to the global model, or the gradients shared during the training process, can infer details about the original data, such as personally identifiable information (PII), even though the raw data was never explicitly shared. This vulnerability arises because the model parameters encode information about the training data, which can be reverse-engineered to reveal private information. To Address and prevent model inversion attacks, we utilise FLTN model parameters such as large batch sizes and normalisation for model training, as well as only sharing local model weights with the community DERMS, and peer-to-peer sharing and augmentation adding an additional layer of privacy.


\subsubsection {\textbf {Model Selection}} \label{sec: Model Selection}

\begin{figure*}[ht]
\centering
\includegraphics[width=15cm, height=5.5cm]{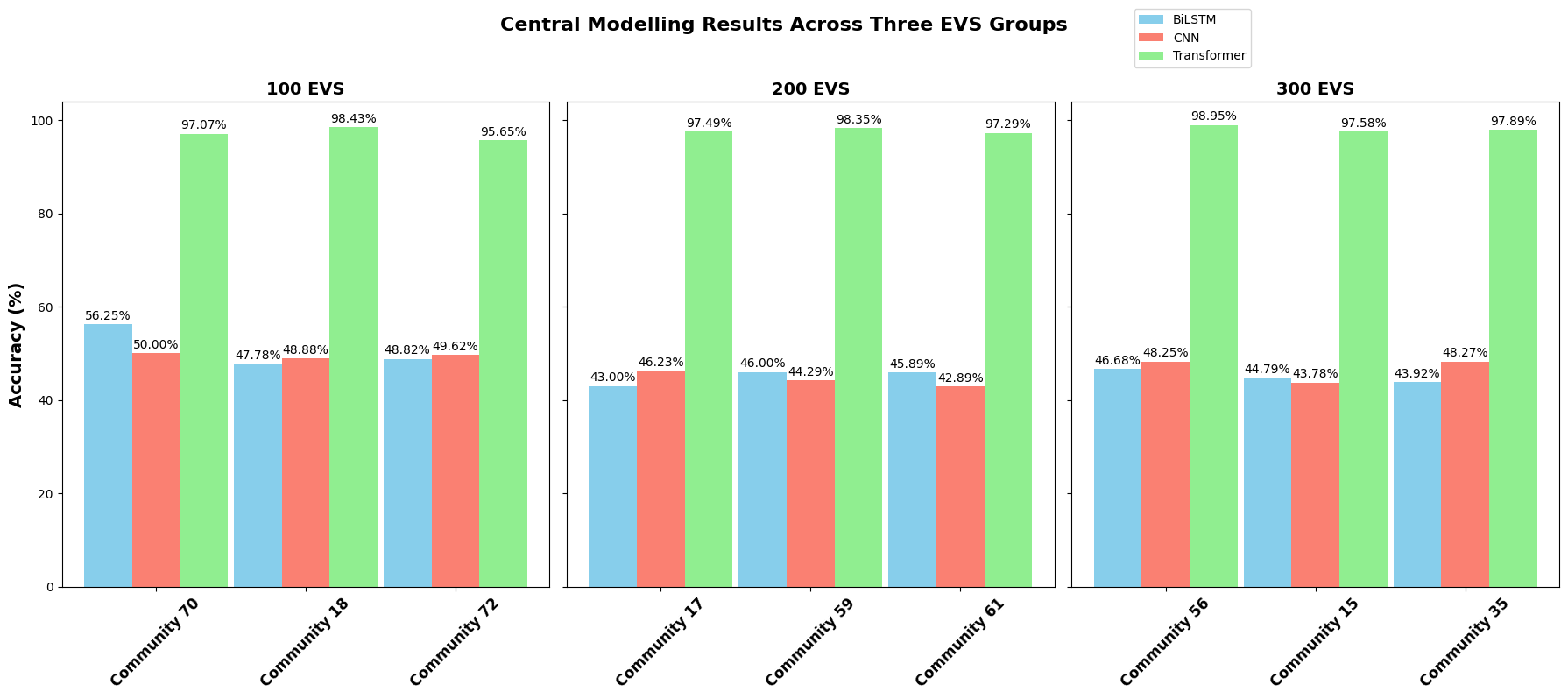}
\vspace{4mm}
\caption{Centralised modelling results. Three machine learning architectures; Transformer, BiLSTM, and CNN. Models used the same communities as our decentralised solution using groups of 100, 200, and 300 EVs to understand which architecture was best suited to our problem area.}
\label{fig: CentralModelling_Results}
\end{figure*}

\par To understand which model would be best suited to our problem area, we experimented with three machine learning models. These models included; a Bi-directional Long Short-Term Memory (BiLSTM) model, then a Convolutional Neural Network (CNN) model, and finally a Transformer model. These models were selected based on their success in previous spatio-temporal mobility research studies \cite{alhudhaif2024spatio} and \cite{elafi2024sta} and \cite{zhang2024research}. 

\par The BiLSTM model was chosen for its proven efficacy in working with spatio-temporal sequences \cite{huang2024spatiotemporal}. BiLSTM’s efficiency in addressing mobility problems stems from its architecture as an advanced Recurrent Neural Network (RNN) that modifies the standard LSTM to process input sequences in both forward and backward directions. This dual processing allows the model to capture context from both past and future states, making it particularly effective for tasks involving sequential data, such as time series prediction, natural language processing, and spatio-temporal analysis. The next model we experimented with consisted of a CNN model. Previous studies by authors Jeon et al.,\cite{jeon2024integrating} and \cite{djenouri2024spatio} discuss the efficiency of spatio-temporal sequence modelling using CNN models. All spatio-temporal sequence data were first converted into images, with each batch represented by images of uniform dimensions. When image sizes vary, the input layer and subsequent filters require substantial preprocessing to handle them, complicating model design and decreasing training efficiency. 

\par Finally, we utilised a PyTorch-based Transformer model, specifically designed to handle our EV mobility spatio-temporal sequence data. This model architecture is particularly well-suited for capturing both spatial dependencies e.g., the relationships between different locations, and temporal dependencies e.g., how these relationships create patterns over time. Our Transformer model was configured with six encoder layers, following standard practices seen in successful implementations such as the original model by Vaswani et al. \cite{vaswani2017attention}. This number of layers was chosen based on the complexity of EV mobility data and the need to capture temporal dependencies. Each encoder layer used eight attention heads, which allow the model to attend to different parts of the input sequence simultaneously, capturing nuanced relationships across time and space. After testing various configurations, this setup provided an optimal balance between computational efficiency and model performance. The multi-head attention mechanism is particularly beneficial for spatio-temporal data, as it allows the model to consider multiple aspects of the sequence concurrently, learning intricate dependencies between spatial regions and time steps.

\subsubsection {\textbf {Training and Validation}}

\par Our solution included machine learning on datasets containing spatio-temporal sequences representing EV taxi trips, with each EV treated as an independent client within our FLTN system. Datasets maintained chronological order of sequences to preserve the temporal dependencies inherent in the data. By treating each EV taxi as a distinct client, we were able to train model weights locally on each EV's dataset before aggregating these weights into a community-level DERMS global model.

\par To ensure a balanced representation of different EV taxi models and to avoid the pitfalls of an imbalanced dataset where some models with higher market shares could dominate the learning process we randomly assigned EV datasets to various taxi vehicles. This method facilitated the preservation of diversity across the FLTN system, allowing our community model to learn from a wide spectrum of driving patterns and charge behaviors unique to each EV model type. The diversity in battery charge levels including 20\%, 40\%, 60\%, 80\%, and 100\% remaining battery charge, covering usage patterns across the different EV taxi models further underscored the importance of maintaining distinct training datasets for each client.

\section {\textbf {Performance Evaluation}} \label{sec: Performance Evaluation}

\par This section compares centralised and decentralised modeling approaches. We begin by examining three centralised models used as baselines to assess their effectiveness in addressing our problem area. Next, we discuss, evaluate, and analyse the final decentralised FL model selected for our proposed solution.
\vspace{5mm}
\subsection {\textbf {Baseline Modelling}} \label{sec: Baseline Modelling}

\par For our baseline modelling experiment we evaluated three centralised models: a Bi-directional LSTM, a CNN, and a transformer model. Our goal was to determine which model could best interpret data representing EV taxi spatio-temporal sequences, encompassing both mobility and charge transaction features. The results of these experiments are presented in the central model results table (Table \ref{tab: CentralModelResultsTable}) and a bar graph (Figure \ref{fig: CentralModelling_Results}). After completing these experiments, it became clear that the transformer model demonstrated the strongest understanding of the problem. Consequently, we selected this model for further decentralised experiments.

\begin{figure}[]
    \includegraphics[width=8.5cm, height=5cm]{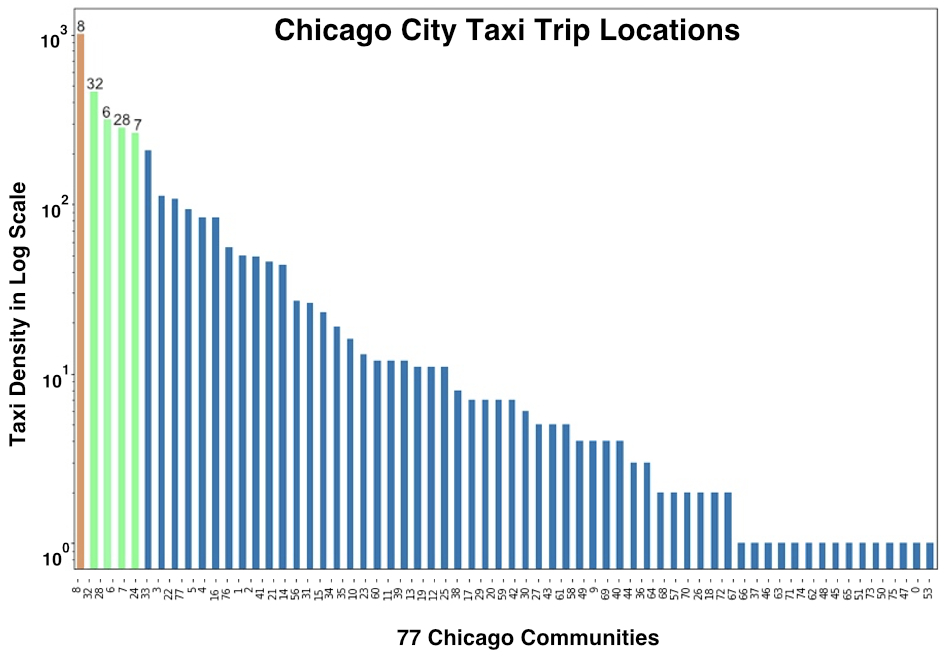}
    \vspace{1mm}
    \caption{This figure represents Chicago city communities and the density of taxi trips per community covering 12 months. The X-axis represents community IDs, the Y-axis density of taxi trips, highest density communities are represented by orange and green, which also represent the CBD for Chicago City.}
    \label{fig: Taxi Density}
\end{figure}
\renewcommand{\arraystretch}{1.2} 
\setlength{\tabcolsep}{8pt}      
\begin{table}[ht]
\centering 
\caption{Central Model Results Table.}
\vspace{4mm}
\label{tab: CentralModelResultsTable}
{\fontsize{7}{8}\selectfont
\begin{tabular}{|l|l|l|l|}
\hline
\textbf{}                               & \textbf{BiLSTM}         & \textbf{CNN}            & \textbf{Transformer}    \\ \hline
\rowcolor[HTML]{EFEFEF} 
{\color[HTML]{333333} \textbf{100 EVS}} & {\color[HTML]{333333} } & {\color[HTML]{333333} } & {\color[HTML]{333333} } \\ \hline
\textbf{Community 70}                   & 56.25\%                 & 50.00\%                 & 97.07\%                 \\ \hline
\textbf{Community 18}                   & 47.78\%                 & 48.88\%                 & 98.43\%                 \\ \hline
\textbf{Community 72}                   & 48.82\%                 & 49.62\%                 & 95.65\%                 \\ \hline
\textbf{}                               &                         &                         &                         \\ \hline
\rowcolor[HTML]{EFEFEF} 
\textbf{200 EVS}                        &                         &                         &                         \\ \hline
\textbf{Community 17}                   & 43.00\%                 & 46.23\%                 & 97.49\%                 \\ \hline
\textbf{Community 59}                   & 46.00\%                 & 44.29\%                 & 98.35\%                 \\ \hline
\textbf{Community 61}                   & 45.89\%                 & 42.89\%                 & 97.29\%                 \\ \hline
\textbf{}                               &                         &                         &                         \\ \hline
\rowcolor[HTML]{EFEFEF} 
\textbf{300 EVS}                        &                         &                         &                         \\ \hline
\textbf{Community 56}                   & 46.68\%                 & 48.25\%                 & 98.95\%                 \\ \hline
\textbf{Community 15}                   & 44.79\%                 & 43.78\%                 & 97.58\%                 \\ \hline
\textbf{Community 35}                   & 43.92\%                 & 48.27\%                 & 97.89\%                 \\ \hline
\end{tabular}\vspace{10pt}\\
}
\footnotesize{Central model results represent the performance of three models: CNN, BiLSTM, and a Transformer. Including EV numbers from 100, 200, and 300,  then separated into 3 communities. Results illustrated by our central modelling plot \ref{fig: CentralModelling_Results}.}
\end{table}

\vspace{5mm}
\subsection {\textbf {Results and Analysis}} \label{sec: Results and Analysis}

\par
\begin{figure*}[h!]
    \centering
    \includegraphics[width=16cm, height=6cm]{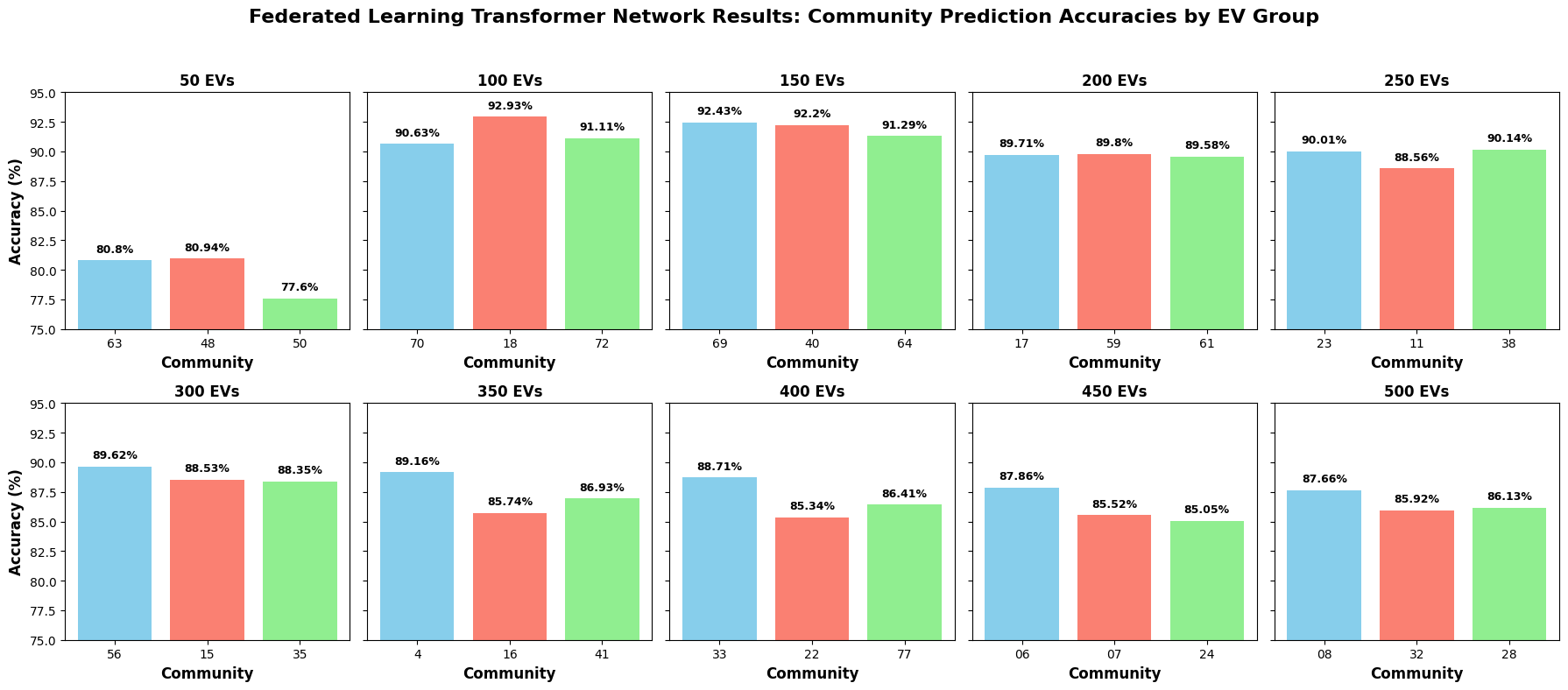}
    \vspace{5mm}
    \caption{The plot above represents 10 EV groups based on EV numbers including 3 community areas each and their respective prediction accuracy for EV next charge location prediction.}
    \label{fig: FLTN Plot}
\end{figure*}

 First we discuss our FLTN results in Table~\ref{tab: FLModelResultsTable} containing the following features, which presents results from an experiment where EVs were split into groups to simulate real-world community structures. Grouping EVs allows us to evaluate the effectiveness of peer-to-peer sharing and augmentation and assess privacy preservation within and across different community models. Our decentalised results table contains the following features \ref{tab: FLModelResultsTable}:

\begin{enumerate}
    \item The first column represents groups of EV numbers ranging from 50 to 500 EVs. These EV densities are based on our taxi density Figure \ref{fig: Taxi Density}. Each EV group contains EVs covering a range of battery charges from 20\% battery remaining through to 100\% battery charge remaining. This use case represents real-world scenarios where communities would include a range of EV model types and battery capacities.    
    \item Each row represents one of the 77 community areas within Chicago, identified by a unique ID number ranging from 1 to 77. These communities, listed in the second column titled 'Total Community List,' were divided into 10 groups based on taxi trip density, as illustrated in Figure~\ref{fig: Taxi Density}. From each group, 3 community areas were selected using simple random sampling, resulting in 10 groups for a total of 30 distinct FLTN models, with related metrics visualised in Figure~\ref{fig: FLTN Plot}
    \item The final columns list the average prediction accuracy and standard deviation for each EV taxi group.  
\end{enumerate}

\par The FLTN table results demonstrate the model's predictive performance across 30 community groups within Chicago city. These communities were selected based on taxi densities as per our taxi density Figure \ref{fig: Taxi Density}. As the number of EVs increases, the model's average prediction accuracy initially rises, peaking within the 100 to 150 EV groups, then gradually declining over subsequent groups. For instance, the 50 EV group achieves an average accuracy of 79.78\%, while the 100 EV and 150 EV groups attain high averages of 91.55\% and 91.97\%, respectively, with low standard deviations (0.9906 and 0.4922). These results suggest that mid-sized EV groups yield more consistent and accurate predictions, likely due to an optimal balance of data diversity and network load. Moreover, this trend indicates that increasing the number of EVs beyond a certain point may introduce additional noise, diminishing the model's predictive capability. The variations in accuracy also highlight the importance of carefully managing data flow within the FLTN framework to maintain performance. 

\par In the larger EV groups, such as those with 200 to 300 EVs, the average prediction accuracy shows a marginal decline, dropping to around 89.69\% for 200 EVs and 88.83\% for 300 EVs. Despite this small decrease, the model maintains reasonable consistency, as shown by the relatively low standard deviations (e.g., 0.0903 for 200 EVs and 0.5610 for 300 EVs). This stability implies that even with increased data volume, the model performs robustly. The largest EV groups (400 to 500 EVs) display a slight decline in accuracy, with averages around 86.82\% to 86.57\% and slightly higher standard deviations (1.4060 for 400 EVs and 0.7754 for 500 EVs). 

\par This pattern suggests that large datasets could introduce multiple EVs that represent 100\% battery charge remaining. These EVs have little to no taxi data which would result in a lower prediction outcome for our model. Larger groups also could reflect diminishing marginal benefit from added data, potentially due to model saturation or the diminishing returns of additional data points within our spatio-temporal sequences. The results align with the study's goals to optimise prediction accuracy across varying data volumes while maintaining privacy, illustrating FLTN's efficacy in balancing performance and scalability in urban EV prediction scenarios.

\renewcommand{\arraystretch}{1.3} 
\setlength{\tabcolsep}{6pt}      
\begin{table*}[t]
\centering
\caption{FLTN Results Table.}
\vspace{3mm}
\label{tab: FLModelResultsTable}
{\fontsize{7}{8}\selectfont
\centering
\begin{tabular}{|
>{\columncolor[HTML]{EFEFEF}}l |l|lll|l|l|}
\hline
\cellcolor[HTML]{ECF4FF}\textbf{No. EVs} & \cellcolor[HTML]{ECF4FF}\textbf{Total Community List}  & \multicolumn{3}{l|}{\cellcolor[HTML]{ECF4FF}\textbf{SRS Communities \& Prediction Accuracies}}               & \cellcolor[HTML]{ECF4FF}\textbf{Average} & \cellcolor[HTML]{ECF4FF}\textbf{Std Dev} \\ \hline
\textbf{50 EVs}                          & 46, 63, 71, 74, 62, 48, 45, 65, 51, 73, 50, 75, 47, 53 & \multicolumn{1}{l|}{\textbf{63:} 80.80\%} & \multicolumn{1}{l|}{\textbf{48:} 80.94\%} & {\textbf{50:} 77.60\%} & 79.78\%                                  & 1.5425                                   \\ \hline
\textbf{100 EVs}                         & 57, 70, 26, 18, 72, 67, 66, 37                         & \multicolumn{1}{l|}{\textbf{70:} 90.63\%} & \multicolumn{1}{l|}{\textbf{18:} 92.93\%} & {\textbf{72:} 91.11\%} & 91.55\%                                  & 0.9906                                   \\ \hline
\textbf{150 EVs}                         & 9, 69, 40, 44, 36, 64, 68                              & \multicolumn{1}{l|}{\textbf{69:} 92.43\%} & \multicolumn{1}{l|}{\textbf{40:} 92.20\%} & {\textbf{64:} 91.29\%} & 91.97\%                                  & 0.4922                                   \\ \hline
\textbf{200 EVs}                         & 17, 29, 20, 59, 42, 30, 27, 43, 61, 58, 49             & \multicolumn{1}{l|}{\textbf{17:} 89.71\%} & \multicolumn{1}{l|}{\textbf{59:} 89.80\%} & {\textbf{61:} 89.58\%} & 89.69\%                                  & 0.0903                                   \\ \hline
\textbf{250 EVs}                         & 10, 23, 60, 11, 39,13, 19, 12, 25, 38                  & \multicolumn{1}{l|}{\textbf{23:} 90.01\%} & \multicolumn{1}{l|}{\textbf{11:} 88.56\%} & {\textbf{38:} 90.14\%} & 89.57\%                                  & 0.7161                                   \\ \hline
\textbf{300 EVs}                         & 14, 56, 31, 15, 34, 35                                 & \multicolumn{1}{l|}{\textbf{56:} 89.62\%} & \multicolumn{1}{l|}{\textbf{15:} 88.53\%} & {\textbf{35:} 88.35\%} & 88.83\%                                  & 0.5610                                   \\ \hline
\textbf{350 EVs}                         & 5, 4, 16, 76, 1, 2, 41, 21                             & \multicolumn{1}{l|}{\textbf{4:} 89.16\%}  & \multicolumn{1}{l|}{\textbf{16:} 85.74\%} & {\textbf{41:} 86.93\%} & 87.27\%                                  & 1.4175                                   \\ \hline
\textbf{400 EVs}                         & 33, 3, 22, 77                                          & \multicolumn{1}{l|}{\textbf{33:} 88.71\%} & \multicolumn{1}{l|}{\textbf{22:} 85.34\%} & {\textbf{77:} 86.41\%} & 86.82\%                                  & 1.4060                                   \\ \hline
\textbf{450 EVs}                         & 6, 7, 24                                               & \multicolumn{1}{l|}{\textbf{06:} 87.86\%} & \multicolumn{1}{l|}{\textbf{07:} 85.52\%} & {\textbf{24:} 85.05\%} & 86.14\%                                   & 1.2289                                   \\ \hline
\textbf{500 EVs}                         & 8, 32, 28                                              & \multicolumn{1}{l|}{\textbf{08:} 87.66\%} & \multicolumn{1}{l|}{\textbf{32:} 85.92\%} & {\textbf{28:} 86.13\%} & 86.57\%                                   & 0.7754                                   \\ \hline
\end{tabular}\vspace{10pt}
}\\
\footnotesize{FLTN Results: The first column represents the number of EVs participating in each row of results, the next column represents all communities within a group based on EV taxi densities derived from our taxi density Figure \ref{fig: Taxi Density}, the next three columns represent a simple random sampling of community areas within Chicago city along with their respective output prediction accuracy. The last two columns represent the average output for this EV group and then our standard deviation for this EV grouping.}
\end{table*}

\renewcommand{\arraystretch}{1.1} 
\setlength{\tabcolsep}{6pt}      
\begin{table}[H]
\centering 
\caption{FLTN Results Table: Over Set Battery Charge Levels.}
\vspace{3mm}
\label{tab: FLTN_SetCharge}
{\fontsize{7}{8}\selectfont
\begin{tabular}{|l|l|l|l|l|l|}
\hline
\textbf{Remaining Capacity} & \textbf{20\%} & \textbf{40\%} & \textbf{60\%} & \textbf{80\%} & \textbf{100\%} \\ \hline
\rowcolor[HTML]{ECF4FF} 
\textbf{100 EV}             &               &               &               &               &                \\ \hline
\textbf{Max}                & 92.14         & 91.89         & 79.33         & 54.73         & 33.33          \\ \hline
\textbf{Min}                & 89.34         & 89.25         & 77.90         & 52.38         & 29.42          \\ \hline
\textbf{Avg}                & 90.74         & 90.57         & 78.61         & 53.55         & 31.37          \\ \hline
\textbf{SD}                 & 1.40          & 1.29          & 0.71          & 1.17          & 1.95           \\ \hline
\rowcolor[HTML]{ECF4FF} 
\textbf{500 EV}             &               &               &               &               &                \\ \hline
\textbf{Max}                & 93.87         & 91.95         & 80.69         & 61.37         & 33.21          \\ \hline
\textbf{Min}                & 91.85         & 90.02         & 78.92         & 59.72         & 30.92          \\ \hline
\textbf{Avg}                & 92.86         & 90.98         & 79.80         & 60.54         & 32.06          \\ \hline
\textbf{SD}                 & 1.01          & 0.96          & 0.88          & 0.82          & 1.14           \\ \hline
\end{tabular}\vspace{10pt}
}\\
\footnotesize{FLTN modelling results for fixed battery charge levels including Max, Min, Avg, and SD for our 100 EV and 500 EV groupings, including 20\%, 40\%, 60\%, 80\% and 100\% battery charge remaining.}
\end{table}

\par Building on our previous experiment with combined battery charge levels Table \ref{tab: FLModelResultsTable}, the FLTN results in Table \ref{tab: FLTN_SetCharge} offer an additional analysis of model performance at distinct charge thresholds. This separation allows us to evaluate the model's predictive accuracy across specific battery levels, adding granularity to the overall findings. In addition, the FLTN set battery charge levels results in Table \ref{tab: FLTN_SetCharge} highlight the model’s predictive capacity across various future states, represented by the 40\%, 60\%, 80\%, and 100\% remaining battery charge columns. These columns reflect the model's effectiveness in forecasting the EVs’ subsequent charging locations over a range of timeframes, spanning from one to three days, depending on the battery capacity of the specific EV model. This predictive capability is grounded in the battery charge remaining, which serves as an indicator of the expected duration of possible taxi trips. 

\par Notably, prediction accuracy is higher at lower battery charge levels due to reduced time (and therefore fewer trips) remaining before the next charging event, thereby minimising uncertainty. Additionally, as the number of accumulated taxi trips increases, the model benefits from a larger dataset, enabling more data-driven and precise forecasting. This relationship underscores the balance between short-term and long-term predictive accuracy: forecasts made closer to the upcoming charge event yield higher precision with a shorter lead time, while those made further in advance, although providing a longer lead time, result in lower accuracy due to increased uncertainty. Analysis suggests an inherent trade-off in predictive performance based on proximity to charge events, highlighting both the model’s robustness in near-term forecasting and the limitations associated with longer prediction horizons. The significance of making predictions for an EV's next charge location over a broad range of battery charge levels using our FLTN method includes several key benefits:

\begin{enumerate}
    \item \textbf{Proactive Charging for EV Owners}: EV owners can rely on this method for timely charging predictions, which helps avoid long queues through proactive management of charging needs.
    \item \textbf{Enhanced Demand Forecasting for Energy Providers}: Energy providers gain insights into energy demand at specific community locations, allowing for proactive adjustments to manage load more effectively and prevent outages.
    \item \textbf{Data Privacy for EV Users}: By protecting EV user data from leakage, the FLTN method prevents malicious actors from inferring private information, safeguarding user privacy.
\end{enumerate}

\par Additionally, trends reveal that as the number of EVs increases, prediction accuracy generally improves. This is because larger numbers of EVs contribute to a more extensive and diverse dataset, allowing the model to capture a broader range of mobility patterns and better distinguish location preferences based on community demand. In contrast, with fewer EVs, the model may experience greater variability and reduced precision due to the limited dataset, especially at higher charge levels where charging events are not immediately required and travel patterns cover greater distances. This trend underscores the importance of dataset size with higher EV counts yielding more stable and generalisable results.
\vspace{2mm}
\subsection{\textbf{Privacy Preservation for Mobility Data}} \label{subsec: Privacy Preservation for Mobility Data}

\par
\begin{figure}[H]
    \centering
    \includegraphics[width=5cm, height=4cm]{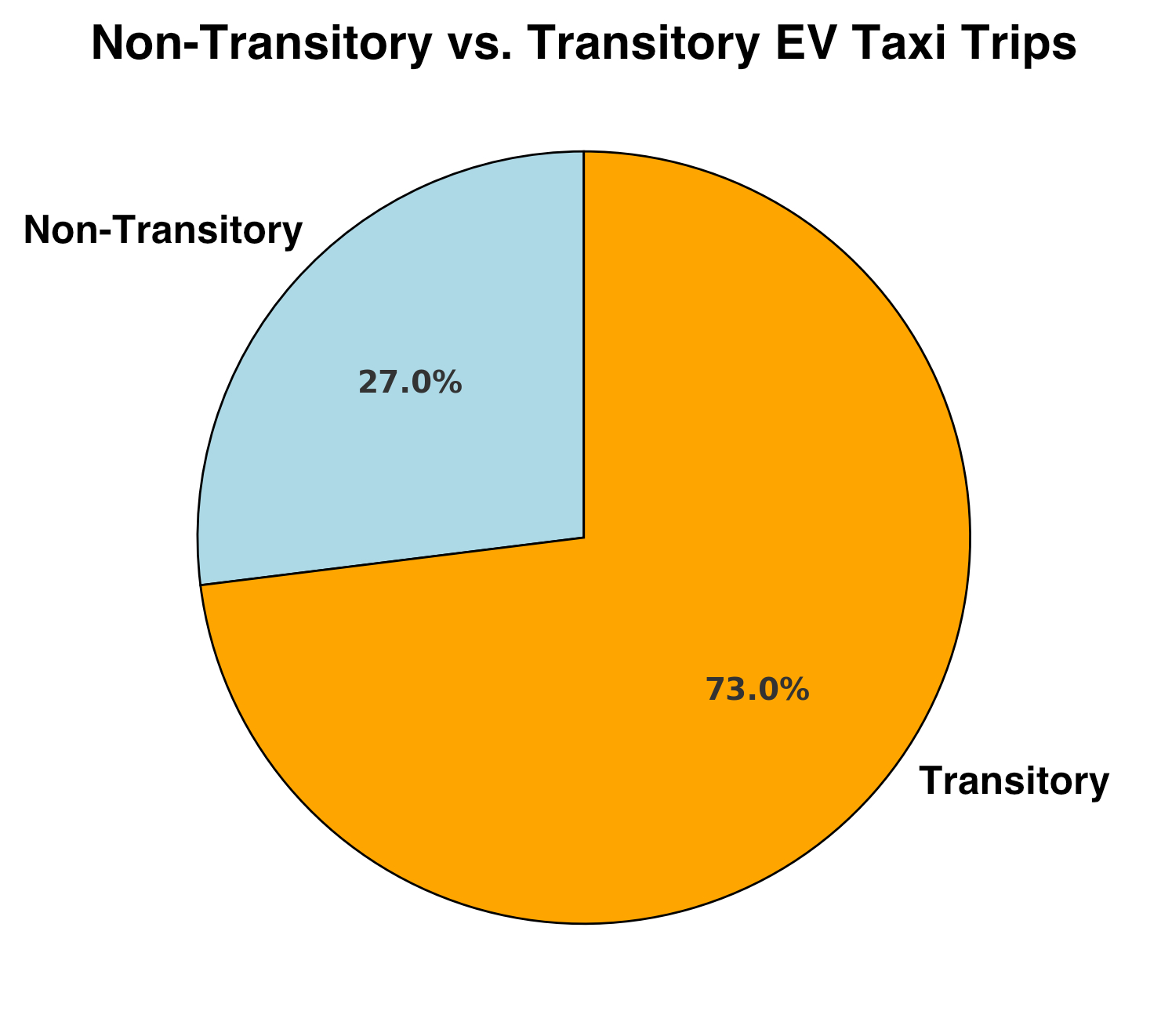}
    \caption{This figure illustrates a sample of 500 EVs over a 12-month period, differentiating transitory and non-transitory EV taxi trips based on pickup and dropoff within the same community.}
    \label{fig: Transitory Plot}
\end{figure}

 This paper introduces a novel solution to protect EV users' private data while maintaining accurate predictions of EVs' next charging locations. The approach leverages spatio-temporal mobility sequences that contain sensitive spatial (location) and temporal (time) information, both of which could reveal individual mobility patterns if not carefully handled. To address privacy risks, this solution differentiates between non-transitory and transitory EVs, reflecting the natural movement patterns of EVs as mobile agents, as shown in Figure~\ref{fig: Transitory Plot}. 

\par Non-transitory EVs, which operate mainly within specific communities, engage in localised peer-to-peer model weight sharing and augmentation before communicating with the community DERMS. This initial exchange helps obfuscate individual EV contributions, adhering to differential privacy principles by making personal data less distinguishable within a group context. On the other hand, transitory EVs, which move across multiple communities, share their model weights directly with the community DERMS, safeguarding broader mobility patterns without compromising privacy. By incorporating peer-to-peer sharing and augmentation, the solution minimises exposure of individual data, promoting both predictive accuracy and enhanced user privacy across varying mobility levels.

\par
\begin{figure}[H]
    \centering
    \includegraphics[width=7.5cm, height=5cm]{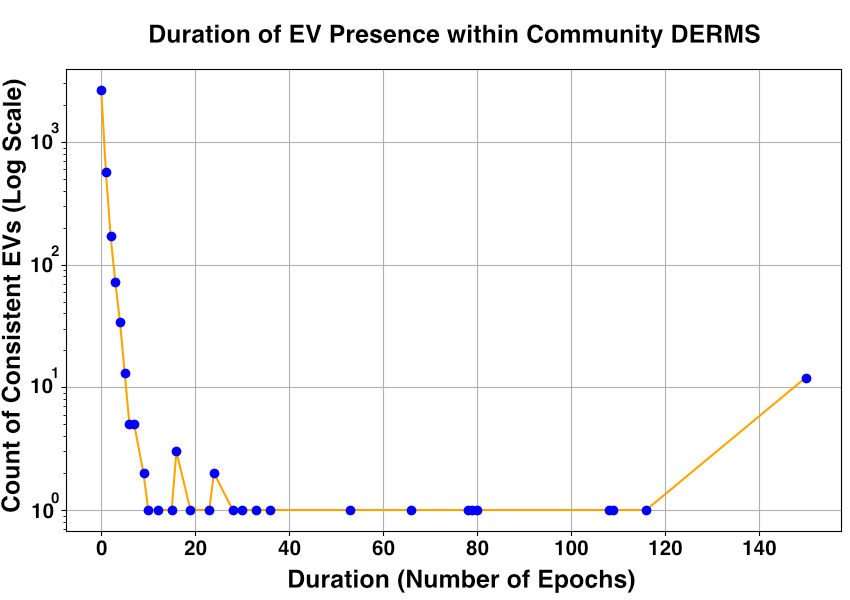}
    \vspace{3mm}
    \caption{Comprehensive view of the EV movement patterns in the DERMS areas.}
    \label{fig: DERMS_mobility}
\end{figure}

 The DERMS mobility Figure \ref{fig: DERMS_mobility} illustrates the rate of EV consistency within a DERMS area, with the X-axis representing the duration (in epochs) an EV stays within the DERMS, and the Y-axis (on a logarithmic scale) showing the count of EVs remaining for each duration. The sharp drop-off at the beginning highlights that most EVs stay for only a few epochs before moving, indicating high mobility within the DERMS area. This behavior aligns with privacy goals in the FLTN framework, as frequent movement makes it challenging for anyone EV to consistently observe others, thus reducing the risk of inference attacks. The logarithmic Y-axis reveals occasional spikes and long-tail behavior at certain durations, suggesting that a few EVs remain longer but are rare. This high turnover rate of EVs, coupled with the 1000 EV sample size, reinforces the robustness of the privacy solution by demonstrating how constantly changing DERMS populations protect against persistent observation attempts.

\par The transitory behavior demonstrated in Figure~\ref{fig: DERMS_mobility} plays a critical role in enhancing privacy by making EV trajectories less predictable. Frequent DERMS changes disrupt potential linkages between specific EVs and consistent spatial patterns, making it harder for adversaries to infer sensitive location data. This dynamic obfuscation improves the anonymisation of shared model updates, thereby reducing the risk of model inversion attacks and strengthening privacy within the FL framework.

\par In addition, the proposed solution incorporates large batch sizes, normalisation, and peer weight sharing and augmentation to further safeguard privacy and model integrity. Sharing model weights instead of gradients minimises vulnerability to reconstruction attacks, as gradients are more prone to revealing sensitive information. Large batch sizes dilute the influence of individual EV data on the DERMS global model, mitigating the potential impact of data poisoning. This balance between predictive accuracy and privacy-preserving measures ensures reliable predictions while protecting sensitive user data in dynamic EV networks.

\subsubsection{Privacy Threat Model}
In the FLTN framework, both DERMS and EVs are considered honest but curious. This implies that while they adhere to the established protocols, they may still attempt to infer additional information from the shared model weights. To mitigate this, the framework incorporates peer weight sharing and augmentation specifically among non-transitory EVs, facilitating weight mixing that obscures direct associations between these EVs and their shared weights. This weight mixing approach prevents DERMS from identifying specific non-transitory EVs through the weights they share, effectively breaking potential links that might compromise privacy. Additionally, the random mobility of transitory EVs adds an inherent layer of anonymity, as their frequent movement across DERMS areas makes it more challenging to associate them with specific weights over time. By anonymising shared weights, the system further reduces the risk of revealing sensitive information, even if DERMS or EVs analyse the data beyond its intended use. This layered approach ensures that privacy is maintained throughout the FL process, minimising potential leakage of sensitive mobility patterns from both DERMS and EVs.


\subsubsection{\textbf{FLTN Privacy Evaluation and Analysis}} \label{subsec: PrivacyAnalysis}

\par To evaluate the privacy efficacy of our FLTN system, we introduce an entropy-based privacy metric, emphasising how increased randomness within shared weights reduces the potential for data reconstruction attacks. Higher entropy values across layers indicate greater uncertainty and obfuscation of individual data, complicating adversarial attempts to infer sensitive information. In FLTN, peer-to-peer sharing and augmentation among non-transitory EVs introduces variability in the shared weights, potentially raising entropy by increasing the uncertainty of the original source. The entropy \( H \) for each layer is calculated as:
\begin{equation}
H = - \sum_{i=1}^n p_i \log(p_i)
\label{eq:entropy}
\end{equation}
where \( p_i \) represents the probability distribution of weights. Unlike regular FL, which aggregates weights uniformly, FLTN’s peer-to-peer sharing and augmentation process introduces variability, potentially yielding higher entropy across critical layers. This increase in entropy enhances privacy by making patterns less traceable.

\par
\begin{figure}[H]
    \centering
    \includegraphics[width=8cm, height=4.5cm]{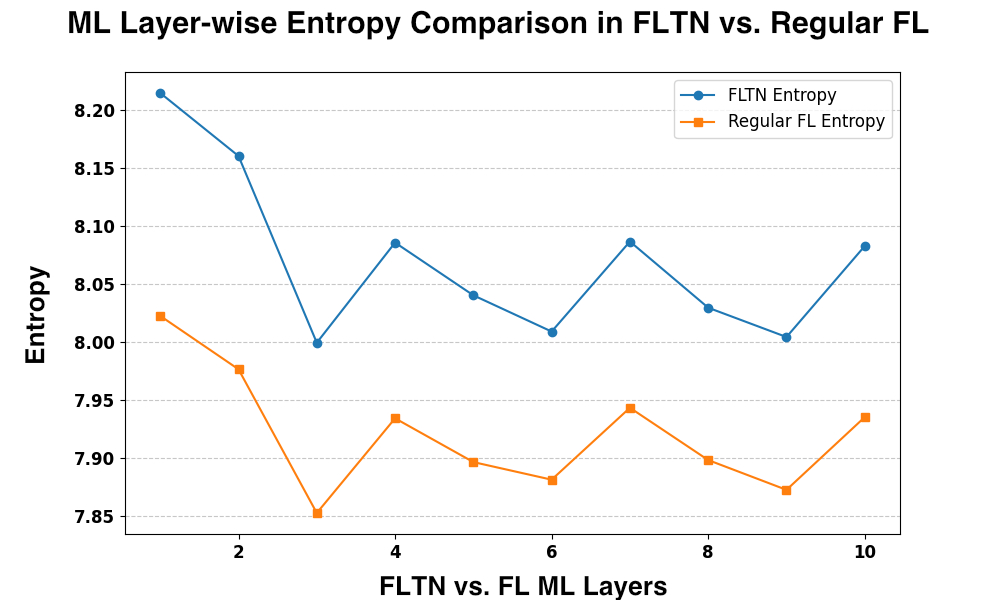}
    \vspace{3mm}
    \caption{Comparison of privacy metrics (entropy) in FLTN vs. regular FL.}
    \label{fig: Privacy_Metrics_Plot}
\end{figure}

 Additionally, FLTN amplifies privacy through minor noise addition within weight-sharing processes. This controlled noise infusion, spread through peer-to-peer exchanges, increases randomness, making it challenging to isolate individual EV patterns or discern specific data trends. The entropy increase of approximately 0.15-0.2 in absolute terms (or 2-2.5\% relative to normal FL entropy) suggests that FLTN provides a more robust privacy layer compared to regular FL. This higher entropy reflects improved obfuscation articulated in Figure \ref{fig: Privacy_Metrics_Plot}, reducing the likelihood of adversaries reconstructing sensitive information from the shared weights. 

\par To further ensure privacy, non-transitory EVs engage in peer-to-peer weight sharing and augmentation before sharing with the community DERMS. Rather than one EV aggregating the weights, this process involves an exchange and mixing of weights among non-transitory EVs, helping to obfuscate individual contributions. By the time the weights reach the DERMS, they have been anonymised through peer-to-peer exchanges, breaking potential links to specific non-transitory EVs. The inclusion of these exchanges and controlled noise addition strengthens FLTN’s privacy, providing greater resistance to data reconstruction attacks compared to regular FL.

\section {\textbf {Discussion and Conclusion}} \label{sec: Discussion}

\par Our proposed FLTN system successfully balances prediction accuracy with enhanced privacy protections, though it introduces some trade-offs inherent to FL approaches in dynamic EV networks. Aggregating model weights across a range of battery levels (20\% to 100\%) improves generalisability and reflects real-world conditions, though this diversity can increase model complexity, potentially impacting training time and computational demands. Large batch sizes, peer-to-peer sharing and augmentation further strengthen privacy by obfuscating individual EV data but may slightly reduce sensitivity to nuanced, individualised patterns, potentially affecting accuracy in unique or rapidly changing scenarios. Nonetheless, FLTN’s multi-layered privacy design offers a robust and scalable solution that not only aligns with privacy regulations but also provides a practical pathway for energy providers to implement data-driven demand forecasting without compromising user privacy. By securely predicting EV charge locations, this system supports privacy-friendly insights into urban mobility patterns, empowering energy providers to meet community energy needs effectively and ethically.

\vspace{5mm}
\subsection*{\textbf{Future Works}} \label{subsec: Future Works}

\par As the landscape of EV mobility evolves, further research is needed to strengthen the security and privacy guarantees of  FLTN. A key area for exploration is the incorporation of advanced privacy-preserving techniques, such as secure multiparty computation and homomorphic encryption. Additionally, embedding real-time anomaly detection mechanisms within the FLTN could help identify and counteract data poisoning attacks more effectively, preserving the integrity of predictive models used in EV infrastructure. Moreover, addressing the trade-offs between security and computational efficiency will be crucial for real-time applications like V2X communication. Future research should also focus on improving model scalability and robustness to ensure the FLTN framework remains resilient in larger and more complex EV networks. By pursuing these advancements, the security and reliability of the EV mobility ecosystem and the power grid can be significantly enhanced.

\vspace{-2.5mm}
\section*{Acknowledgements}
The work has been supported by the Cyber Security Research Centre Limited whose activities are partially funded by the Australian Government’s Cooperative Research Centres Programme.


\bibliographystyle{IEEEtran}
\bibliography{references.bib}

\begin{thebibliography}{10}
\providecommand{\url}[1]{#1}
\csname url@samestyle\endcsname
\providecommand{\newblock}{\relax}
\providecommand{\bibinfo}[2]{#2}
\providecommand{\BIBentrySTDinterwordspacing}{\spaceskip=0pt\relax}
\providecommand{\BIBentryALTinterwordstretchfactor}{4}
\providecommand{\BIBentryALTinterwordspacing}{\spaceskip=\fontdimen2\font plus
\BIBentryALTinterwordstretchfactor\fontdimen3\font minus \fontdimen4\font\relax}
\providecommand{\BIBforeignlanguage}[2]{{%
\expandafter\ifx\csname l@#1\endcsname\relax
\typeout{** WARNING: IEEEtran.bst: No hyphenation pattern has been}%
\typeout{** loaded for the language `#1'. Using the pattern for}%
\typeout{** the default language instead.}%
\else
\language=\csname l@#1\endcsname
\fi
#2}}
\providecommand{\BIBdecl}{\relax}
\BIBdecl

\bibitem{EnerOutlook21}
``The latest update to ener-data’s online,'' \url{https://www.enerdata.net/publications/energy-outlook-tool.html}, 2021, accessed Dec. 3, 2023.

\bibitem{mcmahan2017communication}
B.~McMahan, E.~Moore, D.~Ramage, S.~Hampson, and B.~A. y~Arcas, ``Communication-efficient learning of deep networks from decentralized data,'' in \emph{Artificial intelligence and statistics}.\hskip 1em plus 0.5em minus 0.4em\relax PMLR, 2017, pp. 1273--1282.

\bibitem{regulation2018general}
P.~Regulation, ``General data protection regulation,'' \emph{Intouch}, vol.~25, 2018.

\bibitem{de2018guide}
L.~de~la Torre, ``A guide to the california consumer privacy act of 2018,'' \emph{Available at SSRN 3275571}, 2018.

\bibitem{duan2021ssgd}
J.~Duan, X.~Li, S.~Gao, Z.~Zhong, and J.~Wang, ``Ssgd: A safe and efficient method of gradient descent,'' \emph{Security and Communication Networks}, vol. 2021, no.~1, p. 5404061, 2021.

\bibitem{allahrakha2023balancing}
N.~Allahrakha, ``Balancing cyber-security and privacy: legal and ethical considerations in the digital age,'' \emph{Legal Issues in the digital Age}, no.~2, pp. 78--121, 2023.

\bibitem{andres2013geo}
M.~E. Andr{\'e}s, N.~E. Bordenabe, K.~Chatzikokolakis, and C.~Palamidessi, ``Geo-indistinguishability: Differential privacy for location-based systems,'' in \emph{Proceedings of the 2013 ACM SIGSAC conference on Computer \& communications security}, 2013, pp. 901--914.

\bibitem{gruteser2003anonymous}
M.~Gruteser and D.~Grunwald, ``Anonymous usage of location-based services through spatial and temporal cloaking,'' in \emph{Proceedings of the 1st international conference on Mobile systems, applications and services}, 2003, pp. 31--42.

\bibitem{sweeney2002k}
L.~Sweeney, ``k-anonymity: A model for protecting privacy,'' \emph{International journal of uncertainty, fuzziness and knowledge-based systems}, vol.~10, no.~05, pp. 557--570, 2002.

\bibitem{douaidi2023predicting}
L.~Douaidi, S.-M. Senouci, I.~El~Korbi, and F.~Harrou, ``Predicting electric vehicle charging stations occupancy: a federated deep learning framework,'' in \emph{2023 IEEE 97th Vehicular Technology Conference (VTC2023-Spring)}.\hskip 1em plus 0.5em minus 0.4em\relax IEEE, 2023, pp. 1--5.

\bibitem{teimoori2022secure}
Z.~Teimoori, A.~Yassine, and M.~S. Hossain, ``A secure cloudlet-based charging station recommendation for electric vehicles empowered by federated learning,'' \emph{IEEE Transactions on Industrial Informatics}, vol.~18, no.~9, pp. 6464--6473, 2022.

\bibitem{li2024federated}
Y.~Li, R.~Xie, C.~Li, Y.~Wang, and Z.~Dong, ``Federated graph learning for ev charging demand forecasting with personalization against cyberattacks,'' \emph{arXiv preprint arXiv:2405.00742}, 2024.

\bibitem{warnat2021swarm}
S.~Warnat-Herresthal, H.~Schultze, K.~L. Shastry, S.~Manamohan, S.~Mukherjee, V.~Garg, R.~Sarveswara, K.~H{\"a}ndler, P.~Pickkers, N.~A. Aziz \emph{et~al.}, ``Swarm learning for decentralized and confidential clinical machine learning,'' \emph{Nature}, vol. 594, no. 7862, pp. 265--270, 2021.

\bibitem{zhang2022dim}
Z.~Zhang, Q.~Su, and X.~Sun, ``Dim-krum: Backdoor-resistant federated learning for nlp with dimension-wise krum-based aggregation,'' \emph{arXiv preprint arXiv:2210.06894}, 2022.

\bibitem{zhou2022differentially}
J.~Zhou, N.~Wu, Y.~Wang, S.~Gu, Z.~Cao, X.~Dong, and K.-K.~R. Choo, ``A differentially private federated learning model against poisoning attacks in edge computing,'' \emph{IEEE Transactions on Dependable and Secure Computing}, vol.~20, no.~3, pp. 1941--1958, 2022.

\bibitem{zhu2019deep}
L.~Zhu, Z.~Liu, and S.~Han, ``Deep leakage from gradients,'' \emph{Advances in neural information processing systems}, vol.~32, 2019.

\bibitem{li2024survey}
H.~Li, L.~Ge, and L.~Tian, ``Survey: federated learning data security and privacy-preserving in edge-internet of things,'' \emph{Artificial Intelligence Review}, vol.~57, no.~5, p. 130, 2024.

\bibitem{marlin2024electric}
R.~Marlin, R.~Jurdak, A.~Abuadbba, S.~Ruj, and D.~Miller, ``Electric vehicle next charge location prediction,'' \emph{IEEE Transactions on Intelligent Transportation Systems}, 2024.

\bibitem{chicago2016}
J.~Levy, ``Chicago data portal,'' 2023, \url{https://data.cityofchicago.org/Transportation/Taxi-Trips/wrvz-psew}, Last accessed on 2023-04-20.

\bibitem{batteryhealth}
S.~Blanco, ``How to maximize ev range,'' \url{https://www.jdpower.com/cars/shopping-guides/how-to-maximize-ev-range}, 2022, [Online; accessed 20-March-2023].

\bibitem{ji2024emerging}
S.~Ji, Y.~Tan, T.~Saravirta, Z.~Yang, Y.~Liu, L.~Vasankari, S.~Pan, G.~Long, and A.~Walid, ``Emerging trends in federated learning: From model fusion to federated x learning,'' \emph{International Journal of Machine Learning and Cybernetics}, pp. 1--22, 2024.

\bibitem{bakare2024data}
S.~S. Bakare, A.~O. Adeniyi, C.~U. Akpuokwe, and N.~E. Eneh, ``Data privacy laws and compliance: a comparative review of the eu gdpr and usa regulations,'' \emph{Computer Science \& IT Research Journal}, vol.~5, no.~3, pp. 528--543, 2024.

\bibitem{wu2024application}
X.~Wu, Y.~Wu, X.~Li, Z.~Ye, X.~Gu, Z.~Wu, and Y.~Yang, ``Application of adaptive machine learning systems in heterogeneous data environments,'' \emph{Global Academic Frontiers}, vol.~2, no.~3, pp. 37--50, 2024.

\bibitem{issa2024rve}
W.~Issa, N.~Moustafa, B.~Turnbull, and K.-K.~R. Choo, ``Rve-pfl: Robust variational encoder-based personalised federated learning against model inversion attacks,'' \emph{IEEE Transactions on Information Forensics and Security}, 2024.

\bibitem{alhudhaif2024spatio}
A.~Alhudhaif and K.~Polat, ``Spatio-temporal characterisation and compensation method based on cnn and lstm for residential travel data,'' \emph{PeerJ Computer science}, vol.~10, p. e2035, 2024.

\bibitem{elafi2024sta}
I.~Elafi, N.~Zrira, A.~Kamal-Idrissi, H.~A. Khan, and A.~Ettouhami, ``Sta-sst: Spatio-temporal time series prediction of moroccan sea surface temperature,'' \emph{Journal of Sea Research}, vol. 200, p. 102515, 2024.

\bibitem{zhang2024research}
H.~Zhang, H.~Wang, X.~Zhang, and L.~Gong, ``Research on traffic flow forecasting based on dynamic spatial-temporal transformer,'' \emph{Transportation research record}, vol. 2678, no.~7, pp. 301--313, 2024.

\bibitem{huang2024spatiotemporal}
D.~Huang, J.~He, Y.~Tu, Z.~Ye, and L.~Xie, ``Spatiotemporal information enhanced multi-feature short-term traffic flow prediction,'' \emph{Plos one}, vol.~19, no.~7, p. e0306892, 2024.

\bibitem{jeon2024integrating}
S.~B. Jeon and M.-H. Jeong, ``Integrating spatio-temporal graph convolutional networks with convolutional neural networks for predicting short-term traffic speed in urban road networks.'' \emph{Applied Sciences (2076-3417)}, vol.~14, no.~14, 2024.

\bibitem{djenouri2024spatio}
Y.~Djenouri, A.~N. Belbachir, A.~Cano, and A.~Belhadi, ``Spatio-temporal visual learning for home-based monitoring,'' \emph{Information Fusion}, vol. 101, p. 101984, 2024.

\bibitem{vaswani2017attention}
A.~Vaswani, ``Attention is all you need,'' \emph{Advances in Neural Information Processing Systems}, 2017.

\end{thebibliography}
\begin{IEEEbiography}
[{\includegraphics[width=1in,height=1.25in,clip,keepaspectratio]{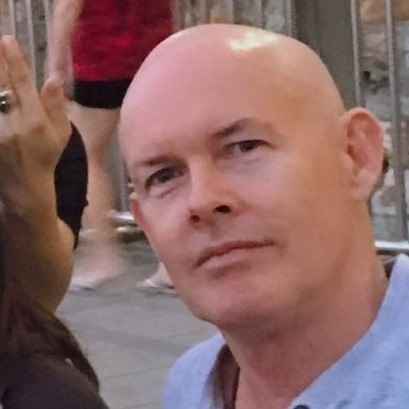}}]
{Robert Marlin}
completed a bachelor’s in information technology (BIT) at the Royal Melbourne Institute of Technology (RMIT University) in 2017, then completed a master’s in information and technology (MIT) from the Queensland University of Technology (QUT) in 2019. He worked as a Research Assistant at QUT in 2019 investigating blockchain within industry applications. Currently, he is undertaking a Ph.D. degree with QUT researching “Distributed and Privacy Preserving Analytics of Smart Grid Data”. He has published a conference paper on Advanced Topics: IoT Camera Vulnerabilities and Weaknesses, as well as co-authored a chapter on Blockchain Model for the Education Sector: A Digital Transformation Perspective of Strategic Learning, for IGI Global Press. Robert's research involves machine learning to resolve issues within the distributed energy domain, including consumer privacy and energy demand prediction.
\end{IEEEbiography}
\vspace{-.5cm} 
\vskip 0pt plus -1fil
\begin{IEEEbiography}[{\includegraphics[width=1in,height=1.25in,clip,keepaspectratio]{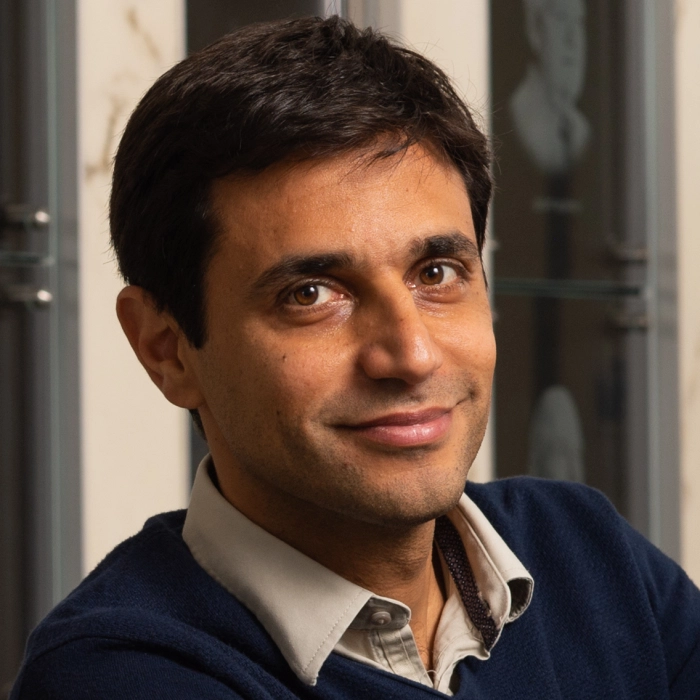}}]{Raja Jurdak}
 is a Professor of Distributed Systems and the Chair of Applied Data Sciences with the Queensland University of Technology, where he is Co-Director of the QUT Energy Transition Centre and Director of the Trusted Networks Lab. He received the M.S. and Ph.D. degrees from the University of California at Irvine. He previously established and led the Distributed Sensing Systems Group, Data61, CSIRO. He also spent time as a Visiting Academic with MIT and Oxford University in 2011 and 2017.  He has published over 250 peer-reviewed publications, including three authored books on IoT, blockchain, and cyberphysical systems. His publications have attracted over 16500 citations, with an H-index of 55. His research interests include trust, mobility, and energy efficiency in networks. He serves on the editorial boards of IEEE Transactions on Network and Service Management and Ad Hoc Networks, and on the organizing and technical program committees of top international conferences, including Percom, ICBC, IPSN, WoWMoM, and ICDCS. He is an IEEE Computer Society Distinguished Visitor, an IEEE Senior Member, and an Adjunct Professor with the University of New South Wales.
\end{IEEEbiography}
\vspace{-.5cm} 
\vskip 0pt plus -1fil
\begin{IEEEbiography}[{\includegraphics[width=1in,height=1.25in,clip,keepaspectratio]{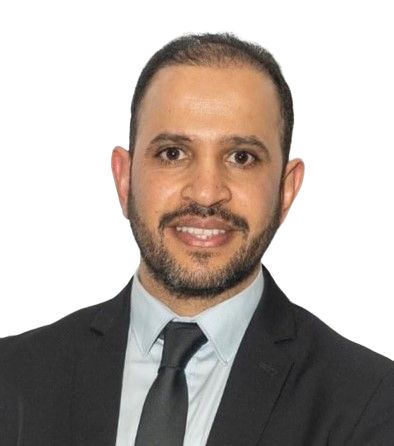}}]{Sharif Abuadbba}
is a team leader of distributed systems security at CSIRO’s Data61 . Sharif has a  Ph.D. in computer security from RMIT  University,  Australia.  He also has several years of experience working as a senior research engineer with Californian-based technology companies. He has contributions to a few US IP filled Patents in cybersecurity. He has 50+ science publications many of which in top venues including IEEE S\&P, NDSS, ACSAC, ASIACCS, ESORICS, IEEE TDSC, and IEEE TIFS. His specialist and interests include AI and cybersecurity, System and data security, and watermarking.
\end{IEEEbiography}
\vspace{-.5cm} 
\vskip 0pt plus -1fil
\begin{IEEEbiography}[{\includegraphics[width=1in,height=1.25in,clip,keepaspectratio]{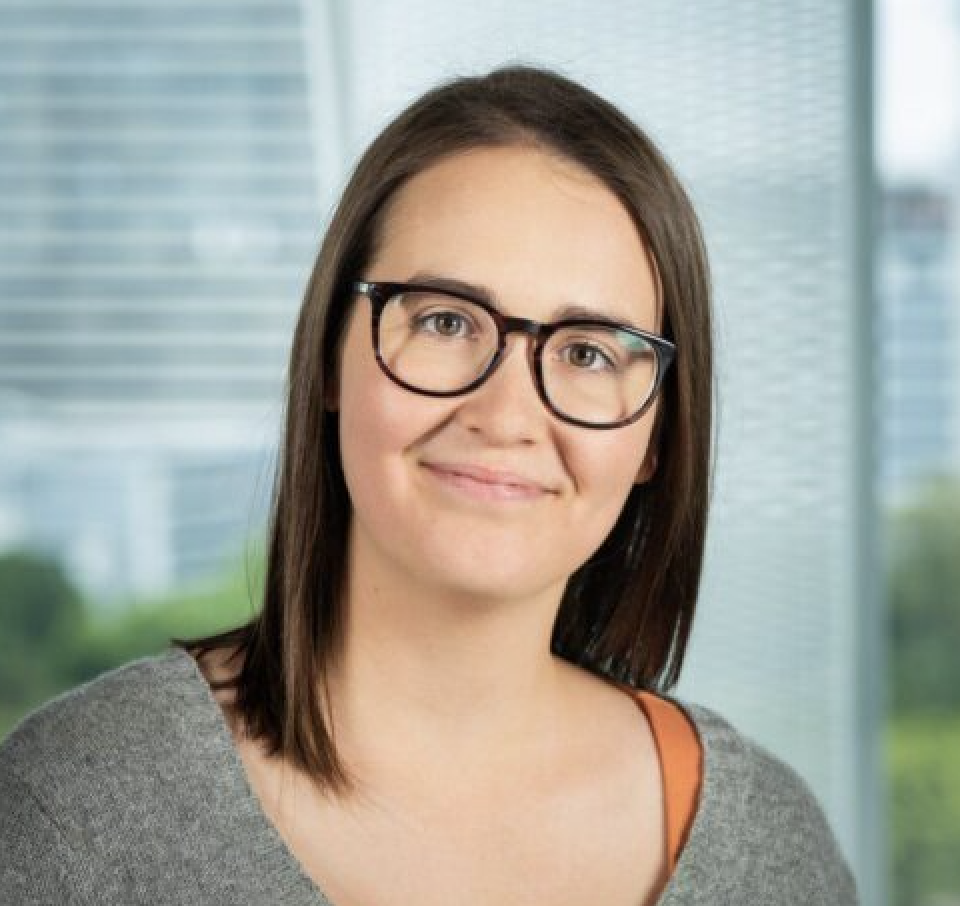}}]{Dimity Miller}
is a Chief Investigator with the QUT Centre for Robotics, and a Lecturer with the QUT School of Electrical Engineering and Robotics. Dimity has a Ph.D. on uncertainty in deep learning for robotic vision, obtained from QUT in 2021 while working within the ARC Centre of Excellence for Robotic Vision. Her thesis was recognised by an Executive Dean’s Commendation for Outstanding Doctoral Thesis Award in 2022. Prior to joining the QUT Centre for Robotics, she worked as a postdoctoral research fellow jointly across the CSIRO Machine Learning and Artificial Intelligence Future Science Platform, the CSIRO Robotics and Autonomous Systems Group, and the QUT Trusted Networks Lab. Dimity's research expertise is in reliable robotic vision – operating at the intersection of deep learning, computer vision, and robotics.
\end{IEEEbiography}

\end{document}